\documentclass [twocolumn,preprintnumbers,amsmath,amssymb]{revtex4}

\usepackage{graphicx}
\usepackage{dcolumn}
\usepackage{bm}

\DeclareGraphicsExtensions{.eps,.gif,.ps}

\begin{document}

\title{Graphical representation of generalized quantum measurements}

\author{Pawe{\l} Kurzy\'nski}

\author{Andrzej Grudka}%

\affiliation{Faculty of Physics,
 Adam Mickiewicz University,
Umultowska 85, 61-614 Pozna\'{n}, Poland.
}

\date{\today}

\begin{abstract}
We present graphical representation for genaralized quantum
measurements (POVM). We represent POVM elements as Bloch vectors and
find the conditions these vectors should satisfy in order to
describe realizable physical measurements. We show how to find
probability of measurement outcome in a graphical way. The whole
formalism is applied to unambigous discrimination of non-orthogonal
quantum states.
\end{abstract}

\maketitle

\newcommand{\ga}{\gamma}
\newcommand{\la}{\lambda}
\newcommand{\ran}{\rangle}
\newcommand{\lan}{\langle}

\section{Introduction}

Quantum measurement is a link between quantum and classical world.
The measurement influence the system that is being measured by
changing it to one of the states that measurement apparatus can
recognize. Neither Schr\"{o}dinger equation, nor any relativistic
description of quantum mechanics, says anything about the dynamics
of measurement. Fortunately, quantum theory can predict the
probabilities of measurement outcomes. For our purpose that is
enough to say something about the physics behind.

Despite the fact that quantum measurement is probabilistic and
"invasive" process it can be found very useful in some interesting
and common (i.e. not only laboratory) applications. For example it
allows us to gain some knowledge on the system being measured. If
someone prepares one of the two known orthogonal quantum states and
asks us what the state is. We can answer his question by performing
appropriate measurement. On the other hand, destructive property of
measurement makes quantum cryptography secure \cite{Bennett}. If
somebody wants to get information about the quantum system he has to
"touch" it and therefore change it. It means that when someone
eavesdrops he leaves "fingerprints" behind him which can be used in
the future to detect his presence.

One often assumes that quantum measurement (the so called von
Neumann measurement) is usually represented by $k \leq N$ operators
where $N=dim({\it{H}})$ is the dimension of the Hilbert space. These
operators satisfy the following coditions:
\begin{eqnarray}
 &A_i \geq 0 \label{e1} \\  &\sum_i A_i = I \label{e2} \\
 &A_iA_j=\delta_{ij}A_i \label{e3}
\end{eqnarray}
and the probability of obtaining during the measurement the result
$i$ is $P(i|\rho)=Tr(A_{i}\rho)$, where $\rho$ is the state of the
quantum system on which the measurement is performed. It has been
shown however, that there are more possible measurement scenarios
than von Neumann orthogonal projective measurements \cite{Peresa,
Helstrom}. This type of measurements are called POVM (Positive
Operator Valued Measure). The three conditions on POVM elements
$A_i$ appeared to be too much. Only two first are necessary to
implement any physical measurement. It may look strange, because we
get use to think of the measurement as of something classical and
when we say classical we usually think of some {\it{orthogonal}}
states. The reason why we can have this counterintuitive operators
is that we can add some ancillary system $B$ then allow it to
interact in a prescribed way with our system $A$ and finally we
perform the collective measurement on both systems \cite{Neumark}.
Getting rid of the last condition gives us a whole bunch of new
possibilities. First of all, we are no longer restricted to only $N$
projective measurement operators. We can have as many operators as
we want as long as they fulfill condition (\ref{e1}) and (\ref{e2}).
The first one simply says that the probability of outcome cannot be
negative. The second one says that the total probability has to sum
up to one. This nonorthogonality of measurement operators allows us
to perform probabilistic unambiguous discrimination of quantum
states - i.e. we can distinguish, with probability less than one,
some quantum states which are in general nonorthogonal
\cite{Ivanovic, Dieks, Peresb, Huttner}.

In this paper we give simple graphical description of POVM
measurements performed on two level quantum systems (qubits) by
representing the POVM elements as vectors inside the Bloch sphere.
We show that using this formalism one can easily calculate the best
strategy for error-free state discrimination of pure states.

\section{POVM elements as sub-normalized quantum states}

Let us consider two restrictions (\ref{e1}) and (\ref{e2}) on POVM
elements. First of all, they have to be positive and therefore
hermitian. As a matter of fact, there are some positive and
hermitian operators that we know very well - density matrices that
describe quantum states. The density matrix posses one feature that
is not necessary for the POVM operators - its diagonal values have
to sum up to one. Taking this into account, we may write any POVM
operator in the following form
\begin{equation} \label{e4}
A_i=a_i\rho_i,
\end{equation}
where $\rho_i$ is some quantum state in the Hilbert space of the
system we want to measure and $a_i$ is real positive constant.

For two level systems one can visualize density matrix as a vector
$\vec{u}$ on or inside the Bloch sphere. We use the standard
formulae \cite{Nielsen} to obtain vector coordinates for the Bloch
vector
\begin{eqnarray}
&u_x=tr(\sigma_x \rho), \label{e5} \\
&u_y=tr(\sigma_y \rho), \label{e6} \\
&u_z=tr(\sigma_z \rho), \label{e7}
\end{eqnarray}
where $\sigma_k$, $k \in \{x,y,z\}$ is the $k$-th Pauli spin matrix.
On the other hand we are able to write any density matrix in the
form
\begin{equation} \label{e8}
\rho=\frac{I}{2}+\frac{\vec{u} \cdot \vec{\sigma}}{2}.
\end{equation}
The vector
$\vec{\sigma}=\vec{x}\sigma_x+\vec{y}\sigma_y+\vec{z}\sigma_z$ where
$\vec{x}$, $\vec{y}$ and $\vec{z}$ are unit vectors.

\begin{figure}
\scalebox{.75} {\includegraphics[width=8truecm]{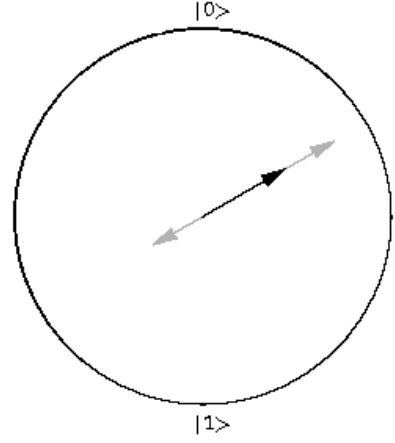}}
\caption{\label{f1} The decomposition of mixed state (black) into a
linear combination of two pure states (gray). It may be written as
$\vec{v_i}=c_{i}\vec{n_{i}}+d_{i}(-\vec{n_{i}})$, where $c_{i}$ and
$d_{i}$ represent the length of gray arrows.}
\end{figure}

At this point we should take a closer look at the density matrix
$\rho_i$ that builds operator $A_i=a_i \rho_i$. If it is a mixed
state we can always decompose it into two (for the two dimensional
Hilbert space) orthogonal pure states. The decomposition goes as
follows: for the mixed state $|\vec{v_i}| < 1$ and therefore
\begin{equation} \label{e9}
\rho_i=\frac{I}{2}+b_{i} \frac { \vec{n_{i}} \cdot \vec{\sigma}}{2},
\end{equation}
where $\vec{v_i}=b_{i} \vec{n_{i}}$ where $b_{i}=|\vec{v_{i}}|$ and
$n_{i}$ is the unit vector. This can be written as
$\rho_i=c_{i}\rho_{i1}+ d_{i}\rho_{i2}$, $1 \geq c_i > d_i \geq 0$
\begin{equation} \label{e10}
\rho_{i1}=\frac{I}{2}+\frac{\vec{n_{i}} \cdot \vec{\sigma}}{2},
\end{equation}
\begin{equation} \label{e11}
\rho_{i2}=\frac{I}{2}-\frac{\vec{n_{i}} \cdot \vec{\sigma}}{2}.
\end{equation}
It is easy to see that $c_{i}+d_{i}=1$ and $c_{i}-d_{i}=b_{i}$. The
graphical representation (see Fig.\ref{f1}) is very simple
\begin{equation} \label{e12}
\vec{v_i}=c_{i}\vec{n_{i}}+d_{i}(-\vec{n_{i}}).
\end{equation}
One should notice that $c_{i}$ and $d_{i}$ are eigenvalues of
$\rho_i$
\begin{equation} \label{e13}
\rho_i = \lambda_{i1} \rho_{i1} + \lambda_{i2} \rho_{i2}.
\end{equation}
But, using this decomposition we split one POVM {\it{mixed}}
operator $A_i$ into two {\it{pure}} operators $A_{i1}$ and $A_{i2}$
\begin{eqnarray} \label{e14}
&A_i=a_i\rho_i = a_i\lambda_{i1} \rho_{i1} + a_i\lambda_{i2}
\rho_{i2}= \nonumber
\\ &a_{i1}\rho_{i1}+a_{i2}\rho_{i2}= A_{i1}+A_{i2}.
\end{eqnarray}
The probability that during the measurement of state $\rho$ the
outcome is $i$ is given by
\begin{equation} \label{e15}
P(i|\rho)=tr(A_i \rho).
\end{equation}
For the mixed operator $A_i$, we see that
\begin{eqnarray} \label{e16}
&P(i|\rho)=tr(A_i \rho)=tr\left((A_{i1}+A_{i2})\rho\right)=\nonumber \\
&=tr(A_{i1}\rho)+tr(A_{i2}\rho)=P(i1|\rho)+P(i2|\rho).
\end{eqnarray}
From now we will consider only operators which are of rank one i.e.
are proportional to pure states. We do not lose generality because
if the operator is of rank two i.e is proportional to mixed state we
can decompose it into two operators of rank one, calculate two
probabilities and then add them. We will visualize POVM as
\begin{equation} \label{e17}
\vec{v_i}=tr(\vec{\sigma}A_i)=a_i tr(\vec{\sigma} \rho_i)=a_{i}
\vec{n_{i}},
\end{equation}
where $\vec{n_{i}}$ is unit vector because $\rho_i$ is pure state,
and quantum states $\rho$ as
\begin{equation} \label{e18}
\vec{r}=tr(\vec{\sigma}\rho).
\end{equation}
Note, that for pure POVM matrix $\rho_i$ the constant $a_i$ plays
the role of the length of $\vec{v_i}$.

Up to now, we have introduced vectors. Right now we will show how to
calculate probabilities of measurement outcome using the scalar
product $\vec{v_i} \cdot \vec{r_j}$. From (\ref{e8}), (\ref{e17})
and (\ref{e18}) we know that
\begin{equation} \label{e19}
A_i=a_i \rho_i=a_i (\frac{I}{2}+\frac{\vec{n_{i}} \cdot
\vec{\sigma}}{2}),
\end{equation}
\begin{equation} \label{e20}
\rho=\frac{I}{2}+\frac{\vec{r} \cdot \vec{\sigma}}{2}.
\end{equation}
Here, it is worth to see the difference between operators $A_i$ and
quantum states. Even if two vectors may look the same in the Bloch
sphere picture, they will look different in the notation (\ref{e19})
and (\ref{e20}). Substituting (\ref{e19}) and (\ref{e20}) into
(\ref{e15}) one obtains
\begin{equation} \label{e21}
P(i|\rho)=\frac{a_i +\vec{v_i}\cdot \vec{r}}{2}.
\end{equation}
because only $I$ and $\sigma_x^2=\sigma_y^2=\sigma_z^2=I$ have
nonzero sum of diagonal elements and contribute to trace.

Moreover, for pure operators $A_i$ and pure states $\rho$ the
probability takes the form
\begin{equation} \label{e22}
P(i|\rho)=\frac{a_i(1 +\cos{\beta})}{2},
\end{equation}
where $\beta$ is the angle between $\vec{v_i}$ and $\vec{r}$.

\begin{figure}
\scalebox{.75} {\includegraphics[width=8truecm]{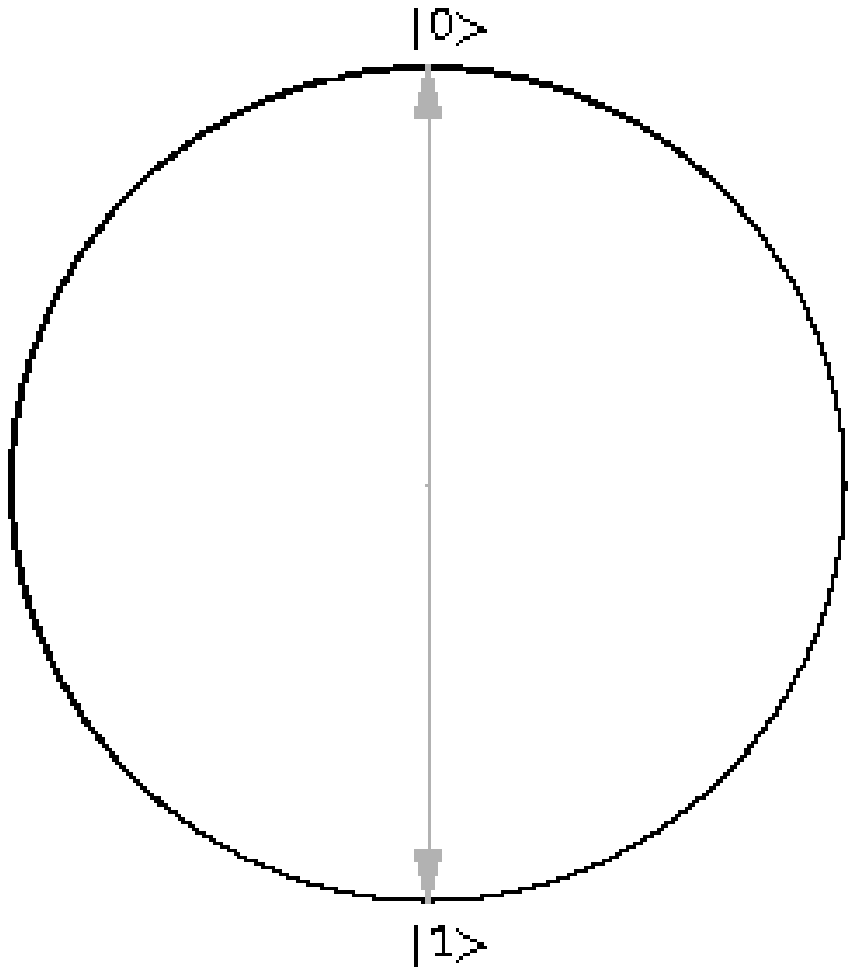}}
\scalebox{.75} {\includegraphics[width=8truecm]{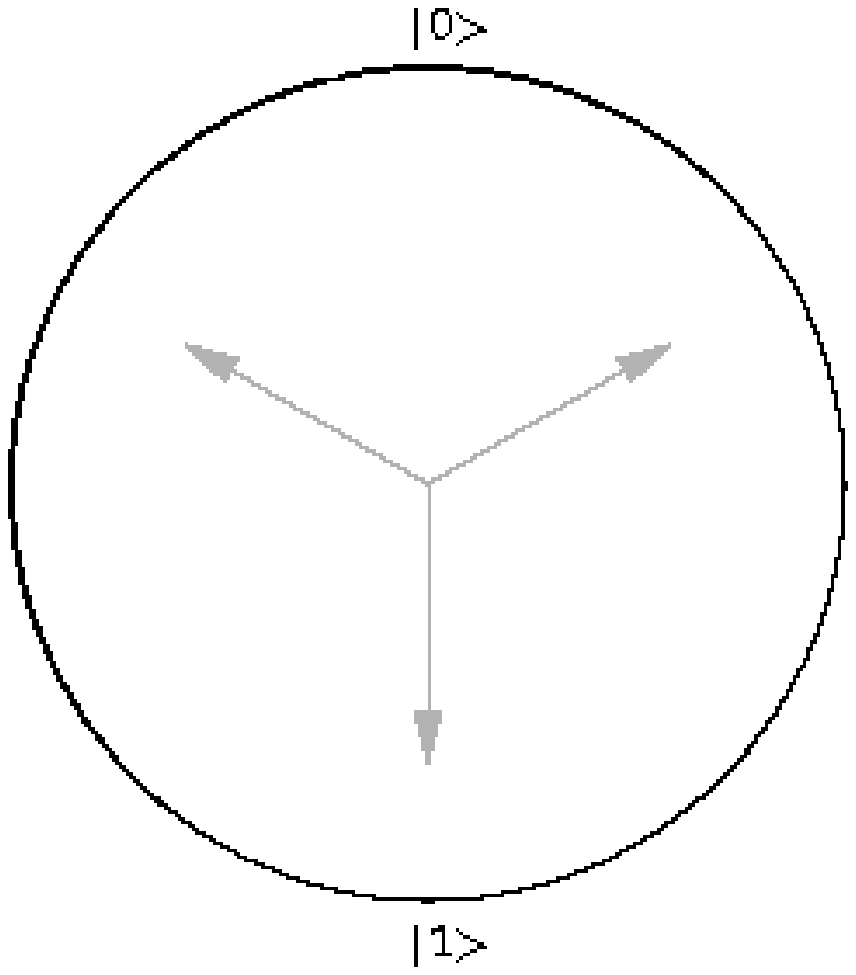}}
\caption{\label{f2} Without loosing generality, we can consider the
Bloch sphere in X-Z plane. Gray arrows represent POVM elements as
vectors. Two arrows (top) are just standard von Neumann projectors.
Three arrows (bottom) represent more general POVM elements. Note,
that the length of the arrows in both cases is different.}
\end{figure}

Next, we will find the restriction on POVM vectors $\vec{v_i}$. If
we sum up all of them and take into account (\ref{e2}) we will
obtain
\begin{equation} \label{e23}
\sum_i \vec{v_i}=\sum_i tr(\vec{\sigma} A_i)=tr(\vec{\sigma} \sum_i
A_i)=tr(\vec{\sigma}I)=0.
\end{equation}
The sum of all vectors has to give zero. This is also visible in
both cases in the Fig.\ref{f2}.

Now, let us say something about $a_i$'s -- the lengths of the
vectors $\vec{v_{i}}$'s. We can derive one condition that they have
to follow by calculating the trace of (\ref{e2}). Trace of the
identity operator is equal to the dimension of the Hilbert space - $
dim({\it{H}})$. The left hand side gives
\begin{equation} \label{e24}
tr(\sum_i A_i)=tr(\sum_i a_i \rho_i) = \sum_i a_i tr(\rho_i)= \sum_i
a_i.
\end{equation}
so our condition for lengths $a_i$'s is
\begin{equation} \label{e25}
\sum_i a_i=dim({\it{H}})=2,
\end{equation}
Therefore we get two conditions for the vectors representing POVM
\begin{eqnarray}
&\sum_i \vec{v_i} = 0, \label{e26} \\
&\sum_i |\vec{v_i}| = 2. \label{e27}
\end{eqnarray}

\section{Error-free discrimination of quantum states}

Imagine that somebody sends us one of two pure states $|\Psi\rangle$
or $|\Phi\rangle$ with the same probability $p=\frac{1}{2}$. We know
what these states are, but we do not know which state the sender has
chosen. Are we able to tell whether we received $|\Psi\rangle$ or
$|\Phi\rangle$? This case was considered earlier by \cite{Ivanovic,
Dieks, Peresb}, but we will solve it in a graphical way. It seems
that if those states are orthogonal we can do it easily, because
this turns out to be purely classical case (even if sometimes the
basis differs from $\{|0\rangle,|1\rangle\}$). We simply perform von
Neumann projective measurement (see Fig.\ref{f2} top).

What happens if $|\Psi\rangle$ and $|\Phi\rangle$ are not
orthogonal? Now, von Neumann measurement is not enough. Although we
might set two measurement vectors $\vec{v_1}$ and $\vec{v_2}$ of
length one ($a_1=a_2=1$) that are parallel (in the Bloch sphere
picture) to one of the states, the other state will be lying on the
Bloch sphere out of the line given by $\vec{v_1}$ and $\vec{v_2}$.
It means, that it may be detected by both $A_1$ and $A_2$ and
therefore, our answers about identity of the state may be wrong. Let
us take $|\Phi\rangle \langle \Phi | = A_1$. In order to be more
convincing we derive the probability of outcome $A_i$ while
measuring the state $|\Psi\rangle$. This probability is given by
\begin{equation} \label{e28}
P(i|\Psi)=tr(|\Psi\rangle\langle\Psi|A_i)=\frac{1}{2}(1+\cos{\beta_i}),
\end{equation}
where $\beta_i$ is the angle between the state vector
$\vec{r_{\Psi}}$ and $\vec{v_i}$ as before. The vector
$\vec{r_{\Psi}}$ is lying out of the line given by $\vec{v_i}$, so
$\beta_i$ is neither zero nor $\pi$ and thus we have that both
probabilities $P(1|\Psi)$ and $P(2|\Psi)$ are nonzero. The
consequences are following. If the outcome of the measurement is
$2$, we know that the only possibility is that the state was
$|\Psi\rangle$. If the outcome is $1$ we know that it probably was
$|\Phi\rangle$, but there is some chance that it might have been
$|\Psi\rangle$, so we are not able to distinguish both states
without an error.

POVM gives us the possibility of error-free discrimination of
quantum states. What we want to obtain is to find two operators
$A_1$ and $A_2$ that have the following property. $A_1$ is
{\it{sometimes}} measured when the first state was sent and is never
measured when the second state was sent. $A_2$ works in the opposite
way. It means, we are looking for the vectors $\vec{v_1}$ and
$\vec{v_2}$ that are antiparallel to the vectors representing the
first and the second state
\begin{eqnarray}
&\vec{v_1} \cdot \vec{r_\Psi}=-|\vec{v_1}||\vec{r_\Psi}|, \label{e29} \\
&\vec{v_2} \cdot \vec{r_\Phi}=-|\vec{v_2}||\vec{r_\Phi}|.~
\label{e30}
\end{eqnarray}
\begin{figure}
\scalebox{.75} {\includegraphics[width=8truecm]{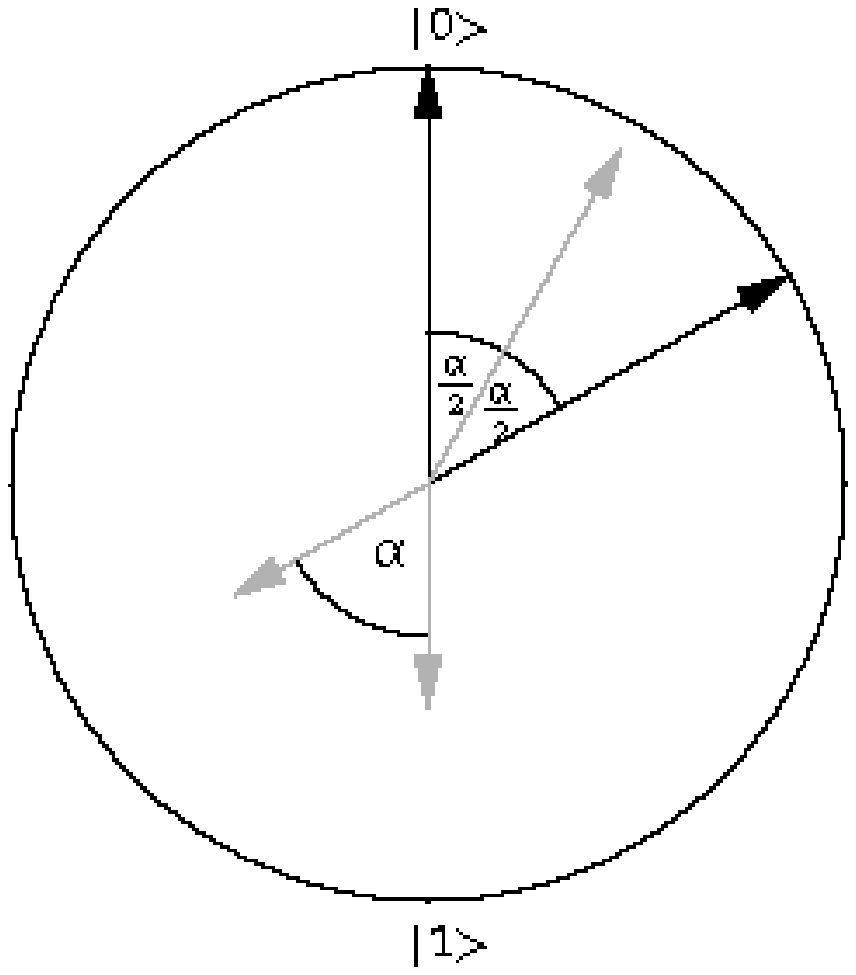}}
\scalebox{.75} {\includegraphics[width=8truecm]{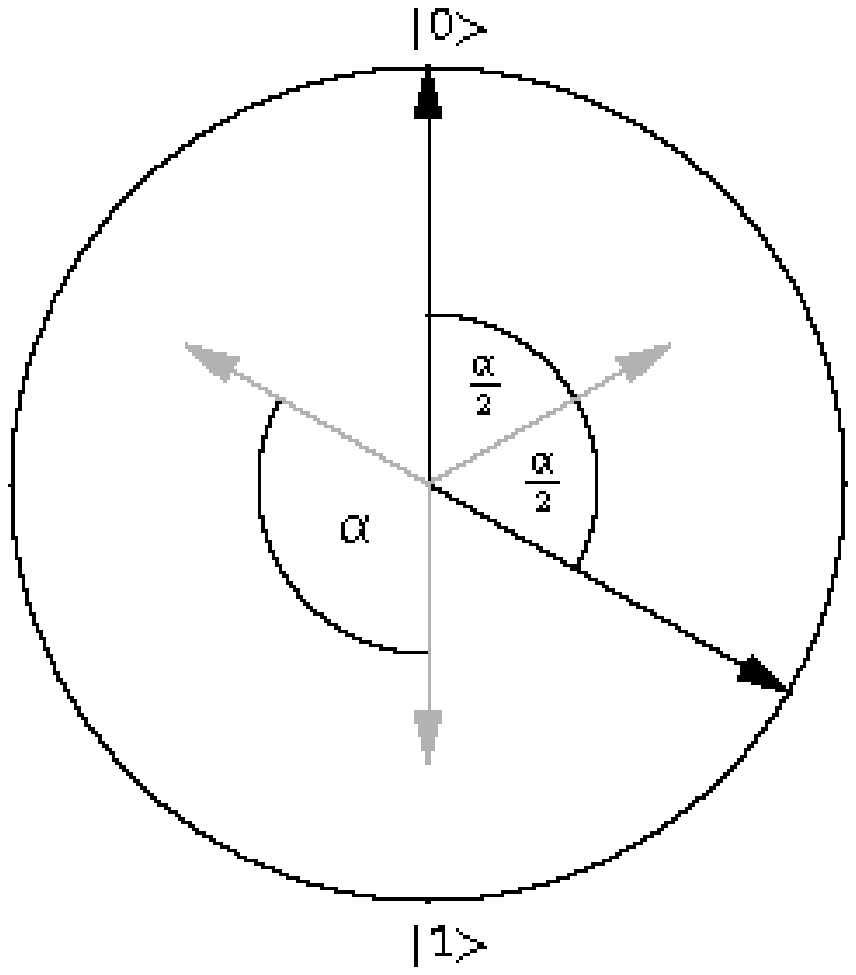}}
\caption{\label{f3} POVM for two different nonorthogonal states.
Black arrows represent two pure states that we want to distinguish
$|0\rangle$ and $|\Psi\rangle=\cos{\frac{\alpha}{2}}|0\rangle +
\sin{\frac{\alpha}{2}}|1\rangle$. Grey arrows represent POVM
elements. The arrow lying exactly in between $|0\rangle$ and
$|\Psi\rangle$ is the error POVM giving no information about which
state was sent. Note, that the more orthogonal are the states we try
to distinguish the shorter the error POVM arrow becomes.}
\end{figure}
\begin{figure}
\scalebox{.75} {\includegraphics[width=8truecm]{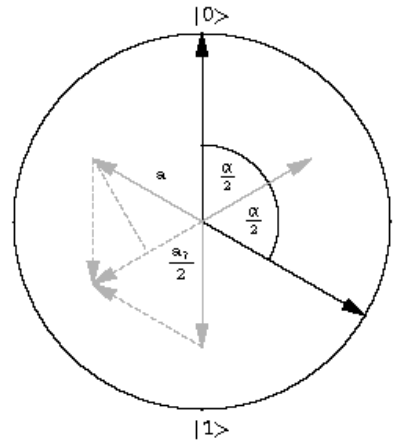}}
\caption{\label{f4} Graphical way to find the direction and the
length $a_?$ of POVM vector $\vec{v_?}.$}
\end{figure}

In order to satisfy (\ref{e26}) we need to introduce the third
vector. We want {\it{"sometimes"}} to happen quite often, that is
why we need $a_1$ an $a_2$ to be as large as possible. Moreover, we
have no bias in favor of any of the states so we put
\begin{equation}\label{e31}
a_1=a_2=a.
\end{equation}
The probability of successful state discrimination is
\begin{equation}\label{e32}
P_{success}=P(1| \rho_{\Phi})=P(2| \rho_{\Psi}),
\end{equation}
where $\rho_{\Phi}=|\Phi\rangle \langle \Phi|$ and
$\rho_{\Psi}=|\Psi\rangle \langle \Psi |$. The equality sign between
two probabilities holds since (\ref{e29}-\ref{e31}) impose symmetry
on states and measurement vectors
$\vec{v_1}\cdot\vec{r_{\Phi}}=\vec{v_2}\cdot\vec{r_{\Psi}}$. Because
of symmetry and condition (\ref{e26}), the third POVM has to point
exactly in between $\vec{r_{\Phi}}$ and $\vec{r_{\Psi}}$ giving
\begin{equation}\label{e33} \vec{v_?}=-(\vec{v_1}+\vec{v_2}).
\end{equation}
Moreover, because of condition (\ref{e27}) we know that
\begin{equation}\label{e34}
a_{?}=2(1-a).
\end{equation}
The reason why we put quotation sign in the subscript of the third
POVM is that it gives completely no information about which state
was sent since
\begin{equation}\label{e35}
\vec{v_?}\cdot\vec{r_{\Phi}}=\vec{v_?}\cdot\vec{r_{\Psi}}.
\end{equation}

Let us write explicit formula for (\ref{e32}) keeping in mind that
$A_i$'s are {\it{pure}} operators and the angle between
$\vec{r_{\Phi}}$ and $\vec{r_{\Psi}}$ is $\alpha$ (see Fig.\ref{f3})
\begin{equation}\label{e36}
P_{success}=\frac{a}{2}(1+\cos(\pi-\alpha))=\frac{a}{2}(1-\cos{\alpha}).
\end{equation}
Then, we go back to (\ref{e33}) and (\ref{e34}) (see also
Fig.\ref{f4}) and obtain
\begin{equation}\label{e37}
a_?=2(1-a)=2a\cos{\frac{\alpha}{2}}.
\end{equation}
We find that
\begin{equation}\label{e38}
a=\frac{1}{1+\cos(\frac{\alpha}{2})}
\end{equation}
and putting this into (\ref{e36}) we get the final formula for the
probability of successful state discrimination
\begin{equation}\label{e39}
P_{success}=\frac{1-\cos(\alpha)}{2(1+\cos(\frac{\alpha}{2}))}.
\end{equation}
Moreover, this is the highest possible probability of success
because we used as few POVM's as possible maximizing their length
$a$.

\section{Conclusions}

We presented simple and intuitive graphical interpretation of
generalized quantum measurements. We have shown that using this
method one can easily calculate the best strategy for error-free
discrimination of two quantum pure states. Although we have not
considered either more than two quantum states or nonuniform
probability distribution of states (i.e. some states are more likely
to appear than other states), we believe that this method could be
applied as well to such cases.

\begin{acknowledgments}
One of us (A. G.) would like to thank the State Committee for
Scientific Research for financial support under Grant No. 0 T00A 003
23.
\end{acknowledgments}

\end{document}